\documentclass[twocolumn,showpacs,pra,floatfix]{revtex4}

\usepackage{graphicx}
\usepackage{bm}
\usepackage{psfrag}
\usepackage{amsmath}
\usepackage{amssymb}
\usepackage{latexsym}
\usepackage{exscale}

\newcommand{\vect}[1]{\bm{#1}}
\newcommand{\ten}[1]{\mbox{\textbf{{\textsf{#1}}}}}

\newcommand{\sprod}{\!\cdot\!}
\newcommand{\tprod}{}

\newcommand{\trace}{\operatorname{Tr}}

\newcommand{\dif}{\mathrm{d}}
\newcommand{\mi}{\mathrm{i}} 
\newcommand{\me}{e}
\newcommand{\be}{\begin{equation}}
\newcommand{\ee}{\end{equation}}

\newcommand{\kB}{k_\mathrm{B}}

\begin{document}

\title{Dynamics of thermal Casimir--Polder forces on polar molecules}

\author{Simen {\AA}dn{\o}y Ellingsen}
\affiliation{Department of Energy and Process Engineering, Norwegian
University of Science and Technology, N-7491 Trondheim, Norway}
\author{Stefan Yoshi Buhmann}
\author{Stefan Scheel}
\affiliation{Quantum Optics and Laser Science, Blackett Laboratory,
Imperial College London, Prince Consort Road,
London SW7 2AZ, United Kingdom}

\date{\today}

\begin{abstract}
We study the influence of thermal Casimir--Polder forces on the
near-surface trapping of cold polar molecules, with emphasis on LiH
and YbF near an Au surface at room temperature. We show that even for
a molecule initially prepared in its electronic and rovibrational
ground state, the Casimir--Polder force oscillates with the
molecule-wall separation. The non-resonant force and the evanescent
part of the resonant force almost exactly cancel at high temperature
which results in a saturation of the (attractive) force in this limit.
This implies that the Casimir-Polder force on a fully thermalised
molecule can differ dramatically from that obtained using a na\"{i}ve
perturbative expansion of the Lifshitz formula based on the molecular
ground-state polarisability. A dynamical calculation reveals how the
spatial oscillations die out on a typical time scale of several
seconds as thermalisation of the molecule with its environment sets
in. 
\end{abstract}

\maketitle


\section{Introduction}
\label{Sec1}

Cold ensembles of polar molecules such as YbF have recently received
particular attention due to their potential use as ultra-sensitive
probes of the permanent electric dipole moment of the electron
\cite{EDM}, measurements of which allow for investigating the possible
existence of physics beyond the standard model \cite{Commins99}. The
need for longer interrogation times has led to the development of
Stark deceleration techniques for these heavy molecules
\cite{MikeYbF,MeijerReview}, with a view to ultimately be able to trap
molecules near microstructured surfaces (chips). Recently, trapping of
light molecules such as metastable CO in travelling potential wells
near a chip surface was achieved \cite{COtrapping}. Another light
diatomic molecule that has received considerable attention due to its
large dipole moment is LiH; and the production of supersonic beams of
cold LiH has been reported \cite{LiHbeam}.

When attempting to trap polar molecules in close proximity to a
surface, attractive Casimir--Polder (CP) forces \cite{Casimir48} ---
effective electromagnetic forces between a neutral and polarisable
particle and a macroscopic object --- need to be taken into account as
an important limiting factor. Thermal CP forces on atoms at thermal
equilibrium with both the electromagnetic field and the present
macroscopic bodies have been intensively studied in the past on the
basis of Lifshitz theory \cite{0057,0057b,0057c,0057d}, linear
response theory \cite{0037,0037b} or normal-mode techniques
\cite{0034,0034b}. At room temperature, the energies associated with
atomic transitions are much larger than the thermal energy,
$\hbar\omega_A\gg k_\mathrm{B}T$, resulting in very low thermal photon
numbers. A ``high-temperature limit'' is only accessible in a
geometric sense when the atom-surface separation $z_A$ is much larger
than the thermal wavelength, $z_A\gg \hbar c/(2\pi k_\mathrm{B}T)$;
in this case the thermal CP force on the atom can be approximated by
\cite{0037,0037b}
\begin{equation}
\label{eq:twolevelatom}
\vect{F}(\vect{r}_A) \approx 
 -\frac{|\vect{d}_A|^2}{8\pi\varepsilon_0z_A^4}
 \,\frac{\kB T}{\hbar\omega_A}
 \,\vect{e}_z
\end{equation}
for a two-level atom (transition frequency $\omega_A$, dipole matrix
element $\vect{d}_A$) interacting with a perfectly conducting plate 
(unit normal $\vect{e}_z$). 

The situation is different for molecules: whereas transition energies
of atoms are typically much larger than attainable thermal energies,
the energies associated with rotational and vibrational transitions of
molecules, heavy molecules in particular, are often small compared to
the thermal energy even at room temperature. A genuine
high-temperature limit $\hbar\omega_A\ll k_\mathrm{B}T$ is hence
realised with an associated large number of thermal photons being
present. An additional consequence of the long transition wavelengths
is the fact that CP forces on molecules are expected to have a long
range with the non-retarded regime $z_A\ll c/\omega_A$ extending quite
far out from the surface. A na\"{i}ve application of the above
formula~(\ref{eq:twolevelatom}) for atoms beyond its scope to the
high-temperature limit $\hbar\omega_A\ll k_\mathrm{B}T$ would suggest
that the force can get arbitrarily strong for molecules of smaller and
smaller transition energies, which already indicates that CP forces on
molecules must be treated with care. 

Supersonic beam expansions typically produce cold molecules that are
to a large fraction in their rovibrational ground states. For example,
in the experiment reported in Ref.~\cite{LiHbeam}, $90\%$ of the
observed cold LiH molecules were in their electronic and rovibrational
ground state $X^1\Sigma^+$. The cold molecule and the room-temperature
surface are thus strongly out of equilibrium with respect to each
other, so a study of the CP interaction necessitates that account be
taken of the full non-equilibrium dynamics of the rotational and
vibrational degrees of freedom of the cold molecule coupled to its
thermal environment. In contrast, in the context of non-equilibrium
forces on thermalised atoms in an environment of non-uniform
temperature, as recently proposed \cite{Antezza05} and measured
\cite{Obrecht07}, a study of the full internal atomic dynamics was not
necessary.

In this article, we study the non-equilibrium thermal CP force on a
polar molecule which is initially in its electronic and rovibrational
ground state in the vicinity of a metal surface. A recently developed
dynamical theory of forces on single atoms or molecules in arbitrary
internal states and arbitrary uniform temperature environments
\cite{ThermoCP} provides the necessary framework (note that a similar
theory has been developed for two-atom van der Waals forces 
\cite{Yury}). In particular, we will show that in contrast to the
above intuitive expectation obtained from comparison with the atom
case, the attractive CP force on a molecule saturates in the
high-temperature limit. 


\section{Casimir--Polder force for given molecular states} 

We consider a polar molecule (energy eigenstates $|n\rangle$,
eigenenergies $\hbar\omega_n$, transition frequencies
$\omega_{mn}\!=\!\omega_m\!-\!\omega_n$, dipole matrix elements
$\vect{d}_{mn}$) which is prepared in an incoherent superposition of
its energy eigenstates with probabilities $p_n$. As shown in
Ref.~\cite{ThermoCP}, the thermal CP force on such a molecule is given
by 
\begin{equation}
\label{eq:totalforce}
\vect{F}(\vect{r}_{\!A})=\sum_{n}p_n\vect{F}_n(\vect{r}_{\!A})
\end{equation}
with perturbative force components
\begin{align}
\label{eq:thermoCP}
&\vect{F}_n(\vect{r}_{\!A})
=-\mu_0k_\mathrm{B}T\sum_{N=0}^\infty
 \bigl(1-{\textstyle\frac{1}{2}}\delta_{N0}\bigr)\xi_N^2
 \nonumber\\ 
&\qquad\times
 \vect{\nabla}_{\!\!A}\trace\bigl[\bm{\alpha}_n(\mi\xi_N) \cdot
 \ten{G}^{(1)}(\vect{r}_{\!A},\vect{r}_{\!A},\mi\xi_N)\bigr]
 \nonumber\\
&\quad+\mu_0\sum_k
\omega_{nk}^2\{
 \Theta(\omega_{nk})[n(\omega_{nk})+1]
 -\Theta(\omega_{kn})n(\omega_{kn})\}\nonumber\\
&\qquad\times\vect{\nabla}_{\!\!A}\vect{d}_{nk}\sprod\operatorname{Re}
 \ten{G}^{(1)}(\vect{r}_{\!A},\vect{r}_{\!A},|\omega_{nk}|)
 \sprod\vect{d}_{kn}
\end{align}
and molecular polarisability
\begin{equation}
\label{eq:polarizability}
\bm{\alpha}_n(\omega)
=\lim_{\epsilon\to 0}
\frac{1}{\hbar}\sum_k\biggl[
 \frac{\vect{d}_{kn}\tprod\vect{d}_{nk}}
{\omega+\omega_{kn}+\mi\epsilon}
-\frac{\vect{d}_{nk}\tprod\vect{d}_{kn}}
 {\omega-\omega_{kn}+\mi\epsilon}
 \biggr].
\end{equation}
Here, $\ten{G}^{(1)}$ is the scattering part of the classical Green
tensor for the electromagnetic field in the given environment and
$\xi_N=2\pi k_\mathrm{B}T N/\hbar$ denotes the Matsubara frequencies.
The CP force~(\ref{eq:thermoCP}) contains both non-resonant
contributions (first term) and resonant ones (second term), where the
former would also follow from applying Lifshitz theory in conjunction
with the ground-state polarisability (we refer to it as Lifshitz-like
force in the following) and the latter are due to the absorption and
emission of thermal photons with photon number
\be
  n(\omega) =
 \frac{1}{e^{\hbar\omega/(k_\mathrm{B}T)}-1}\,.
\ee
Given a probability distribution $p_n$, Eq.~(\ref{eq:thermoCP}) allows
us to compute the thermal CP force. In particular, if the molecule is
in an isotropic state such as the ground state or a thermal state (see
below), the forces simplifies to \cite{ThermoCP}
\begin{align}
\label{eq:thermoCPiso}
&\vect{F}_n(\vect{r}_{\!A})
=-\mu_0k_\mathrm{B}T\sum_{N=0}^\infty
 \bigl(1-{\textstyle\frac{1}{2}}\delta_{N0}\bigr)
 \xi_N^2\alpha_n(\mi\xi_N)\nonumber\\ 
&\qquad\times
 \vect{\nabla}_{\!\!A}\trace\bigl[
 \ten{G}^{(1)}(\vect{r}_{\!A},\vect{r}_{\!A},\mi\xi_N)\bigr]
 \nonumber\\
&\quad+\frac{\mu_0}{3}\sum_k
\omega_{nk}^2\{
 \Theta(\omega_{nk})[n(\omega_{nk})+1]
 -\Theta(\omega_{kn})n(\omega_{kn})\}\nonumber\\
&\qquad\times|\vect{d}_{nk}|^2\vect{\nabla}_{\!\!A}
 \trace\operatorname{Re}
 \ten{G}^{(1)}(\vect{r}_{\!A},\vect{r}_{\!A},|\omega_{nk}|)
\end{align}
with
\begin{equation}
\label{eq:polarizabilityiso}
\alpha_n(\omega)
=\lim_{\epsilon\to 0}
\frac{1}{3\hbar}\sum_k\biggl[
 \frac{|\vect{d}_{nk}|^2}{\omega+\omega_{kn}+\mi\epsilon}
-\frac{|\vect{d}_{nk}|^2}{\omega-\omega_{kn}+\mi\epsilon}
 \biggr].
\end{equation}


\subsection{Molecule near a plane surface}

For example, let us consider a molecule at a distance $z_A$ from the
planar surface of a (non-magnetic) substrate. The respective
scattering Green tensor is given by \cite{Tomas95}
\begin{multline}
\label{eq:planargreen}
\ten{G}^{(1)}(\vect{r},\vect{r},\omega)
=\frac{\mi}{8\pi}\int _0^\infty \dif q\, \frac{q}{\beta}\,
e^{2i\beta z}\\
\times\biggl[\biggl( r_s
-\frac{\beta^2c^2}{\omega^2}\,r_p\biggr)
(\vect{e}_x\vect{e}_x+\vect{e}_y\vect{e}_y)
+2\,\frac{q^2c^2}{\omega^2}\,r_p
\vect{e}_z\vect{e}_z\biggr],
\end{multline}
where
\be
  r_s=\frac{\beta-\beta_1}{\beta+\beta_1}, 
 ~~~ r_p=\frac{\varepsilon(\omega)\beta-\beta_1}
 {\varepsilon(\omega)\beta+\beta_1}
\ee
with $\mathrm{Im}\,\beta$, $\mathrm{Im}\,\beta_1\geq 0$ are the
Fresnel reflection coefficients for $s$- and $p$-polarised waves, 
$\beta=\sqrt{\omega^2/c^2-q^2}$ and 
$\beta_1=\sqrt{\varepsilon(\omega)\omega^2/c^2-q^2}$ are the
$z$-components of the wave vectors in free space and inside the
substrate, and $\varepsilon(\omega)$ is the (relative) permittivity of
the substrate. Substitution of
$\ten{G}^{(1)}(\vect{r},\vect{r},\omega)$ into
Eq.~(\ref{eq:thermoCPiso}) above leads to an explicit form for the CP
force. 

The results simplify in the non-retarded and retarded limits of small
and large atom-surface separations. In the non-retarded limit
$\max_i\{|\sqrt{\varepsilon(\omega_i)}|
\omega_i\} z_A/c\ll 1$ ($\omega_i$: relevant molecular and medium
frequencies), the approximation $\beta\simeq\beta_1\simeq\mi q$ leads
to 
\begin{align}
\label{eq:thermoCPnr}
&\vect{F}_n(\vect{r}_{\!A})
=-\frac{3k_\mathrm{B}T}{8\pi\varepsilon_0z_A^4}
 \sum_{N=0}^\infty
 \bigl(1\!-\!{\textstyle\frac{1}{2}}\delta_{N0}\bigr)
 \alpha_n(\mi\xi_N)\,
 \frac{\varepsilon(\mi\xi_N)-1}{\varepsilon(\mi\xi_N)+1}\, 
 \vect{e}_z\nonumber\\
&-\frac{1}{8\pi\varepsilon_0z_A^4}\sum_k|\vect{d}_{nk}|^2
 \biggl\{\Theta(\omega_{nk})[n(\omega_{nk})\!+\!1]\,
 \frac{|\varepsilon(\omega_{nk})|^2-1}
 {|\varepsilon(\omega_{nk})+1|^2}\nonumber\\
&\quad -\Theta(\omega_{kn})n(\omega_{kn})\,
 \frac{|\varepsilon(\omega_{kn})|^2-1}
 {|\varepsilon(\omega_{kn})+1|^2}\biggr\} 
 \vect{e}_z.
\end{align}
Note that while applying well to dielectrics, the non-retarded limit
often provides a very poor approximation for metals, because the large
factor $|\sqrt{\varepsilon(\omega_i)}|$ may restrict its range of
applicability to extremely small distances.

In the retarded limit $\omega_\mathrm{min}z_A/c\gg 1$
($\omega_\mathrm{min}$: minimum of the relevant molecular and medium
frequencies), the resonant part of the force is well approximated by
letting $q\simeq 0$, while the approximations
$\alpha_n(\mi\xi_N)\simeq\alpha_n(0)$ and
$\varepsilon(\mi\xi_N)\simeq\varepsilon(0)$ hold for those $\xi_N$
giving the main contribution to the non-resonant part. The
$q$-integral for the $N=1$ term can then be carried out immediately,
while those for the remaining part of the sum can be rewritten in a
more convenient form by introducing the integration variable $v=\beta
c/\xi_N$. Performing the sum according to
\begin{equation}
\label{eq:sum}
\sum_{N=1}^\infty N^4y^N=\frac{y^4+11y^3+11y^2+1}{(1-y)^5}\,,
\end{equation}
one finds
\begin{align}
\label{eq:thermoCPret}
&\vect{F}_n(\vect{r}_{\!A})
=-\frac{3k_\mathrm{B}T\alpha_n(0)}{16\pi\varepsilon_0z_A^4}
 \frac{\varepsilon(0)-1}{\varepsilon(0)+1}\, 
 \vect{e}_z\nonumber\\
&\quad-\frac{k_\mathrm{B}T\alpha_n(0)}{2\pi\varepsilon_0z_A^4}
 \int_1^\infty\dif v\,v\,
 \biggl[-\frac{v-\sqrt{\varepsilon(0)-1+v^2}}
 {v+\sqrt{\varepsilon(0)-1+v^2}}\nonumber\\
&\qquad+(2v^2-1)\frac{\varepsilon(0)v-\sqrt{\varepsilon(0)-1+v^2}}
 {\varepsilon(0)v+\sqrt{\varepsilon(0)-1+v^2}}\biggr]\nonumber\\
&\qquad\times
 \frac{x^4(\me^{-8vx}+11\me^{-6vx}+11\me^{-4vx}+\me^{-2vx})}
 {(1-\me^{-2vx})^5}\,\vect{e}_z\nonumber\\
&\quad+\frac{\mu_0}{6\pi cz_A}\sum_k|\vect{d}_{nk}|^2
 \biggl\{\Theta(\omega_{nk})\omega_{nk}^3[n(\omega_{nk})+1]\nonumber\\
&\qquad\times
 \operatorname{Im}\biggl[\me^{2\mi\omega_{nk}z_A/c}
 \frac{\sqrt{\varepsilon(\omega_{nk})}-1}
 {\sqrt{\varepsilon(\omega_{nk})}+1}\biggr]\vect{e}_z
 -\Theta(\omega_{kn})\omega_{kn}^3\nonumber\\
&\qquad\times n(\omega_{kn})
 \operatorname{Im}\biggl[\me^{2\mi\omega_{kn}z_A/c}
 \frac{\sqrt{\varepsilon(\omega_{kn})}-1}
 {\sqrt{\varepsilon(\omega_{kn})}+1}\biggr]\vect{e}_z
 \biggr\}
\end{align}
where $x=2\pi k_\mathrm{B}Tz_A/(\hbar c)$. In particular, for a
conductor whose plasma frequency $\omega_\mathrm{P}$ is large compared
to $\omega_{nk}$ [cf.~Eq.~(\ref{Drude}) below] one has
$|\varepsilon|\gg 1$ and the retarded CP force is well approximated by
\begin{align}
\label{eq:thermoCPretpc}
&\vect{F}_n(\vect{r}_{\!A})
\approx-\frac{3k_\mathrm{B}T\alpha_n(0)}{16\pi\varepsilon_0z_A^4}\, 
 \vect{e}_z
 -\frac{k_\mathrm{B}T\alpha_n(0)}{8\pi\varepsilon_0z_A^4}\,
 \frac{1}{(\me^{2x}-1)^4}\nonumber\\
&\qquad\times\bigl[\bigl(3+6x+6x^2+4x^3\bigr)\me^{6x}
 -\bigl(9+12x-16x^3\bigr)\me^{4x}\nonumber\\
&\qquad+\bigl(9+6x-6x^2+4x^3\bigr)\me^{2x}
 -3\bigr]\vect{e}_z\nonumber\\
&\quad+\frac{\mu_0}{6\pi cz_A}\sum_k|\vect{d}_{nk}|^2
 \biggl\{\Theta(\omega_{nk})\omega_{nk}^3[n(\omega_{nk})+1]\nonumber\\
&\qquad\times\sin(2\omega_{nk}z_A/c)\vect{e}_z
 -\Theta(\omega_{kn})\omega_{kn}^3n(\omega_{kn})\nonumber\\
&\qquad\times \sin(2\omega_{kn}z_A/c)\vect{e}_z
 \bigr\}.
\end{align}
Note that the retarded limit as given above holds for all distances
which are sufficiently large with respect to the atomic and medium
wavelengths, irrespective of the temperature. When in addition the
distance is very large with respect to the thermal wavelength (such
that $x\gg 1$), the contribution from the second terms in the above
Eqs.~(\ref{eq:thermoCPret}) and (\ref{eq:thermoCPretpc}) vanishes and
the non-resonant force approaches its well-known (geometric)
high-temperature limit, cf.~Eq.~(\ref{eq:twolevelatom}). In the
opposite case of a distance which is much smaller than the thermal
wavelength ($x\ll 1$), the first terms vanish and the non-resonant
force reduces to its (retarded) zero-temperature form,
cf.~Ref.~\cite{0019}. Our results, in particular those for the
non-retarded limit, agree with the ones previously obtained in
Ref.~\cite{0034b}. Note that resonant force components and their
oscillatory behaviour in the retarded regime were first discussed for
excited atoms at zero temperature (cf., e.g., \cite{0292,0042}).

The limits reveal that the CP force follows a $1/z_A^4$ power law for
non-retarded distances. In the retarded regime, the non-resonant force
components again follow an inverse power law whereas the resonant
force components give rise to spatially oscillating forces whose
amplitude is proportional to $1/z_A$. If present, the resonant force
components are dominating over the non-resonant ones, in general. The
magnitude of the contributions from various molecular transitions to
the force~(\ref{eq:thermoCP}) is determined by their dipole matrix
elements and frequencies, where Eqs.~(\ref{eq:thermoCPnr}) and
(\ref{eq:thermoCPret}) together with Eq.~(\ref{eq:polarizabilityiso})
imply that the strength of the non-resonant force is roughly
proportional to $1/\omega_{kn}$, while that of the resonant force is
governed by $n(\omega_{nk})+1$ or $n(\omega_{kn})$ in the non-retarded
limit and by $\omega_{nk}^3[n(\omega_{nk})+1]$ or
$\omega_{kn}^3n(\omega_{kn})$ in the retarded regime.
Equations~(\ref{eq:thermoCPnr}) and (\ref{eq:thermoCPret}) furthermore
show that the force becomes larger for larger permittivity of the
surface material and saturates in the high-conductivity limit.

The general results and discussion given above can be easily applied
to various polar molecules interacting with different surface
materials. The qualitative behaviour of the forces will be similar for
all molecules and materials, i.e., a power-law dependence for
non-retarded distances will give way to an oscillating force in the
retarded regime. The exact magnitude of the force as well as the
length scale of the oscillations will depend on the dipole moments
and frequencies associated with the specific molecular transitions
involved and the electric response of the surface in the way indicated
above. Tabulated data for a variety of molecules and metal surfaces
can be found in Ref.~\cite{Heating}. In the following, we will
consider two representative examples. 


\subsection{Examples: LiH and YbF near an Au surface}

\begin{figure}[t]
\includegraphics[width=\columnwidth]{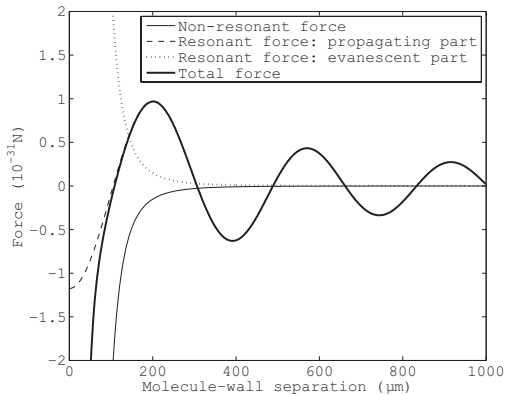}
\caption{\label{fig:LiHgroundstate} Thermal CP force on a
ground-state LiH molecule near an Au surface. For details, see text.}
\end{figure}

We first consider a LiH molecule in its electronic, vibrational,
and rotational ground state ($p_n=\delta_{n0}$) near an Au surface
at room temperature $T=300$~K. With the help of the Green tensor
(\ref{eq:planargreen}), we are able to compute the force components
according to Eq.~(\ref{eq:thermoCPiso}) which are displayed in
Fig.~\ref{fig:LiHgroundstate}. For this molecule, the contribution
from rotational transitions with
$\omega_{kn}=2.79\times 10^{12}\,\mathrm{rad}/\mathrm{s}$ and
$d=1.96\times10^{-29}\,\mathrm{Cm}$ \cite{Heating}
($\sum_k\vect{d}_{0k}\vect{d}_{k0}=d^2\ten{I}$ where
$\ten{I}$ is the unit tensor) is strongly dominant over
those of vibrational and electronic transitions with their
considerably higher transition frequencies. Molecules with a similar
behaviour include NH, OH, OD, NaCs and KCs. For the relative
permittivity of the Au surface we have used a Drude model
\be
\label{Drude}
 \varepsilon(\omega)=
 1-\frac{\omega_\mathrm{P}^2}{\omega(\omega+\mi\gamma)}
\ee
with $\omega_\mathrm{P}=1.37\times 10^{16}\,\mathrm{rad}/\mathrm{s}$
and $\gamma=5.32\times 10^{13}\,\mathrm{rad}/\mathrm{s}$
\cite{lambrecht00}. In view of the current debate regarding the
thermal Casimir force (cf.\ \cite{brevik06,klimchitskaya06} and
references therein), we have also calculated the force using the
alternative plasma model and found that the difference between the two
models is of no importance in our case. 

Figure~\ref{fig:LiHgroundstate} shows the contributions from the
non-reso\-nant force component (thin solid line) which is seen to be
strictly attractive, and the resonant force components (dashed and
dotted lines). With regard to the latter, we have separately shown the
propagating part ($q\in[0,\omega/c)$ in Eq.~(\ref{eq:planargreen}),
dashed line) and the evanescent part ($q\in[\omega/c,\infty)$, dotted
line). The rather astonishing result is that the evanescent part of
the resonant force almost exactly cancels the non-resonant force
component. Hence, in this highly non-equilibrium situation the largest
contribution to the CP force arises from the propagating part of the
resonant force. The total force (thick solid line in
Fig.~\ref{fig:LiHgroundstate}) thus closely follows the latter. Only
at very small molecule-wall separation $z$ the force is given by its
near-field part which, for a two-level isotropic molecule with
$\hbar\omega_A\ll k_BT$, reads
\begin{eqnarray}
\label{eq:twolevel}
\vect{F}(\vect{r}_A) &\!\!=&\!\!
\frac{|\vect{d}_A|^2}{8\pi\varepsilon_0z_A^4}
\left[n(\omega_A)
\frac{|\varepsilon(\omega_A)|^2\!-\!1}
{|\varepsilon(\omega_A)\!+\!1|^2}
-\frac{k_BT}{\hbar\omega_A}\,
\frac{\varepsilon(0)\!-\!1}{\varepsilon(0)\!+\!1}
 \right]\vect{e}_z  \nonumber \\ 
&\!\!\approx&\!\! 
\frac{|\vect{d}_A|^2}{8\pi\varepsilon_0z_A^4}
\left[n(\omega_A) -\frac{k_BT}{\hbar\omega_A} \right]\vect{e}_z \,.
\end{eqnarray}
The approximation in the second line of Eq.~(\ref{eq:twolevel}) holds
for good conductors. The force saturates in the high-temperature limit
where the factor in square brackets approaches $-1/2$. In contrast,
the non-resonant (Lifshitz-like) force alone would formally diverge.
The predicted high-temperature saturation agrees with the previously
found vanishing of the leading, linear contribution in
$k_BT/(\hbar\omega_A)$ in the good-conductor limit \cite{0034b}.

Let us next consider a molecule that is at thermal equilibrium with
its environment, so that the probabilities $p_n$ are given by a
Boltzmann distribution,
\be
 p_n=\frac{\me^{-\hbar\omega_n/(k_BT)}}
 {\sum_j\me^{-\hbar\omega_j/(k_BT)}}.
\ee
Here, all resonant force components cancel and the force is given by
single a non-resonant force contribution given by the first term in
Eq.~(\ref{eq:thermoCPiso}) where the molecular polarisability has to
be replaced by its thermal counterpart \cite{ThermoCP}
\begin{equation}
\alpha_T(\omega)=\sum_np_n\alpha_n(\omega).
\end{equation}
\begin{figure}[t]
\includegraphics[width=2.8 in]{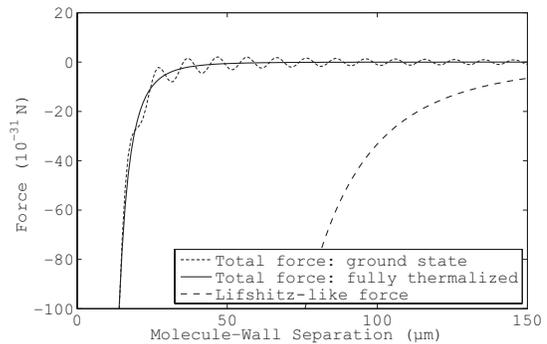}
\caption{\label{fig:YbF} Thermal CP force on a ground-state vs fully
thermalised YbF molecule near an Au surface. For details, see text.}
\end{figure}
In Fig.~\ref{fig:YbF}, we compare this equilibrium force on a
thermalised molecule (solid line) with the non-equilibrium
ground-state force (dotted line) for the case of YbF.
In contrast to LiH, both rotational ($\omega_{kn}=9.05\times
10^{10}\,\mathrm{rad}/\mathrm{s}$,
$d=1.31\times10^{-29}\,\mathrm{Cm}$)
and vibrational transitions
($\omega_{kn}=9.54\times 10^{13}\,\mathrm{rad}/\mathrm{s}$,
$d=8.60\times 10^{-31}\,\mathrm{Cm}$)
\cite{Heating} give relevant contributions to the force, 
because at room temperature the frequency of the latter is very close
to the peak of the spectrum $\omega_{kn}^3n(\omega_{kn})$ determining
the resonant force contributions in the retarded limit. The results
for YbF are thus representative of those to be expected for CaF, BaF,
LiRb, NaRb, LiCs, which also have considerable contributions from
vibrational transitions at room temperature. Figure~\ref{fig:YbF}
shows that in contrast to the ground-state force, which oscillates
as a function of molecule-wall separation (due to the influence of
vibrational transitions), the force on a fully thermalised atom is 
monotonous and attractive (dominated by rotational transitions). 
We emphasise that the force at thermal equilibrium between the atom
and its environment (solid line) is vastly overestimated by a
Lifshitz-type macroscopic calculation (dashed line) that uses the
ground-state polarisability $\alpha_0(\omega)$ as input parameter. The
reduction factor in the near-field limit is approximately given by
\cite{ThermoCP} as
\be
  \frac{|\vect{F}|}{|\vect{F}_\mathrm{Lifshitz}|}
 \simeq \frac1{2n(\omega_{10})+1}
\ee
for all $\vect{r}_A$. Its dependence on the relevant transition
frequency clearly makes it species-dependent. The potentially very
large reduction factors ($\approx 1/870$ for YbF at room temperature)
imply that these molecules can be brought much closer to metallic
surfaces than previously thought. 


\section{Dynamical Casimir--Polder force}

In order to understand the transition between the non-equilibrium
ground-state force and the fully thermalised one, we need to
investigate the full internal molecular dynamics in the presence of
the Au surface. The time-dependent probabilities $p_n=p_n(t)$ are
governed by the rate equations
\begin{equation}
\dot{p}_n(t) =-\sum_k \Gamma_{nk} p_n(t)
 +\sum_k\Gamma_{kn} p_k(t),
\end{equation}
where the transition rates are given by \cite{Heating}
\begin{multline}
\Gamma_{nk}
=\frac{2\mu_0}{\hbar}\omega_{nk}^2
 \left\{\Theta(\omega_{nk})[n(\omega_{nk})+1]
+\Theta(\omega_{kn})n(\omega_{kn})\right\}\\
\times \vect{d}_{nk}\sprod\operatorname{Im}
 \ten{G}(\vect{r}_{\!A},\vect{r}_{\!A},|\omega_{nk}|)
 \sprod\vect{d}_{kn} \,.
\end{multline}

\begin{figure}[t]
\includegraphics[width=3.3in]{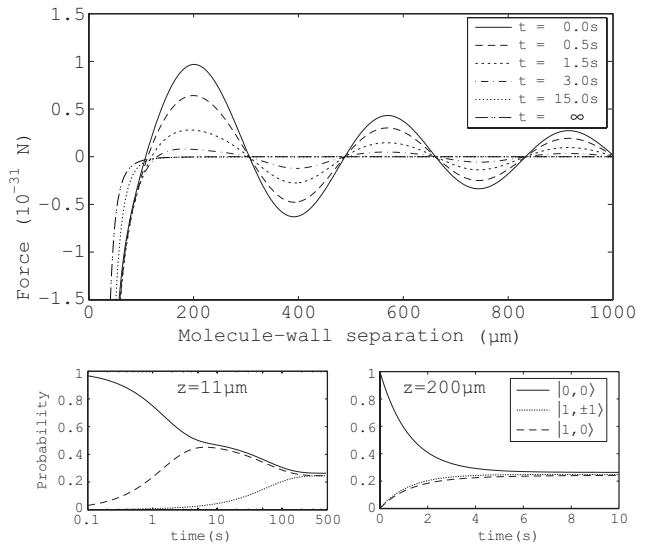}
\caption{\label{fig:transition} Transient CP force and
internal dynamics of a LiH molecule initially prepared in its ground
state. For details, see text.}
\end{figure}

The transition rates for LiH near an Au surface can easily be
calculated using the Green tensor (\ref{eq:planargreen}). The
resulting time-dependent probabilities $p_n(t)$ are displayed for
the ground state and the first manifold of rotationally excited states
in the lower panels in Fig.~\ref{fig:transition}, with the respective
transition matrix elements being given by
$\vect{d}_{|0,0\rangle\to|1,M\rangle}=d\vect{u}_M$,
$\vect{u}_{0}=\vect{e}_z/\sqrt{3}$, 
$\vect{u}_{\pm 1}=(\mp\vect{e}_x+\mi\vect{e}_y)/\sqrt{6}$
\cite{Heating}. For large molecule-wall separation the transition
rates to the different substates of the first manifold are very
similar and so are the resulting probabilities (lower right panel).
When moving closer to the surface, the transition rates become
affected by the evanescent and propagating parts of the reflected
field. The contributions of the latter are strongly oscillating so
that the rates $\Gamma_{|0,0\rangle\to|1,\pm 1\rangle}$ exhibit a
pronounced minimum at $z=11\mu$m. This is not the case for the rate
$\Gamma_{|0,0\rangle\to|1,0\rangle}$ due to the $1/z$-contribution
from the evanescent fields (lower left panel in
Fig.~\ref{fig:transition}). Hence, at first only the occupation of the
level $|1,0\rangle$ reaches equilibrium with the level $|0,0\rangle$,
and full thermalisation is realised only at a much later time.

The dynamics of the CP force is then governed by the internal
molecular dynamics according to
\begin{equation}
\label{eq:ThermoCPtime}
\vect{F}(\vect{r}_{\!A},t)=\sum_{n}p_n(t)\vect{F}_n(\vect{r}_{\!A}).
\end{equation}
This time-dependent force is shown for a LiH molecule initially
prepared in its ground state $p_n(t\!=\!0)=\delta_{n0}$ in the top
panel of Fig.~\ref{fig:transition}. We observe a gradual disappearance
of the oscillating force components on a timescale of approximately
$3$s. The attractive near-field force reaches its equilibrium value
only much later due to the above mentioned strongly reduced rate
$\Gamma_{|0,0\rangle\to|1,\pm 1\rangle}$. Note that during
the thermalisation the molecule is in an anisotropic state 
so that we have to use the general expression~(\ref{eq:thermoCP}) for
the force components rather than its isotropic special
case~(\ref{eq:thermoCPiso}).


\section{Conclusions and outlook}

Studying the CP force on polar molecules near a planar surface at
finite temperature, we have found that even ground-state molecules are
subject to resonant spatially oscillating force components at finite
temperature. They are due to the thermal non-equilibrium between the
molecule and its environment. A full dynamical treatment has shown
that these transient forces disappear in the course of thermalisation
of the molecule. The remaining equilibrium force can be vastly
different from that calculated using a Lifshitz-type force expression
for ground-state molecules. 

In our numerical example of ground-state LiH, we have explicitly shown
that the non-resonant force component and the evanescent part of the
resonant force component cancel almost exactly, leaving a strongly
reduced attractive force in the non-retarded limit which saturates at
high temperatures. The force in the retarded limit is dominated by
resonant contributions from rotational transitions. In contrast, the
force on the heavier molecule YbF is dominated by resonant
contributions from vibrational transitions. Moreover, in thermal
equilibrium at room temperature the resulting force is a factor
$1/870$ smaller than would be expected from a Lifshitz-type
calculation for the corresponding ground-state molecule.

Whereas the CP force on a fully thermalised molecule is always
attractive, the non-equilibrium force on a ground-state molecule near
an Au surface at room temperature has been found to show an
oscillating behaviour as a function of the molecule-wall separation
$z_A$, with stable equilibrium positions away from the surface.
Therefore, one might be tempted to use these (transient) minima for
trapping purposes. It turns out, however, that for LiH the first
potential well (with its minimum at $z_A=300\mu$m) has a depth of
approximately $10^{-12}$K which is immeasurably small. In order to
increase the trap depth, one might envisage a situation in which the
molecule is embedded in a planar cavity of size $l$ consisting of two
such Au surfaces. Then, the Fresnel reflection coefficients in
Eq.~(\ref{eq:planargreen}) have to be replaced by 
$\tilde{r}_{s,p} = r_{s,p}/(1-r_{s,p}^2 e^{2\mi\beta l}).$
For very good conductors such as Au, one can set
$|r_{s,p}|\approx 1-\eta$ with $\eta\ll 1$. Hence, for
$\beta l=n\pi$ ($n\in\mathbb{N}$), the modified Fresnel coefficients
increase as $|\tilde{r}_{s,p}|\propto 2/\eta$. 
Choosing $l$ close to a cavity resonance $n\pi c/\omega_A$, the
contribution from propagating modes with small $q$ can thus be boosted
by several orders of magnitude. Thus, if by use of a cavity with a
high $Q$-factor could increase the trap depth by a factor e.g.\ $10^6$, 
the energy difference would be in the microkelvin regime which could
be sufficiently deep to trap cold polar molecules with thermal
photons. This question will be addressed in more detail in a future
investigation.


\acknowledgments
This work was supported by the Alexander von Humboldt Foundation and
the UK Engineering and Physical Sciences Research Council. S.Y.B. is
grateful to M.R. Tarbutt for discussions.


\end{document}